 \documentclass[sigconf]{acmart} 

\AtBeginDocument{%
  \providecommand\BibTeX{{%
    \normalfont B\kern-0.5em{\scshape i\kern-0.25em b}\kern-0.8em\TeX}}}

\copyrightyear{2023}
\acmYear{2023}
\setcopyright{acmlicensed}\acmConference[CHI '23]{Proceedings of the 2023 CHI Conference on Human Factors in Computing Systems}{April 23--28, 2023}{Hamburg, Germany}
\acmBooktitle{Proceedings of the 2023 CHI Conference on Human Factors in Computing Systems (CHI '23), April 23--28, 2023, Hamburg, Germany}
\acmPrice{15.00}
\acmDOI{10.1145/3544548.3581020}
\acmISBN{978-1-4503-9421-5/23/04}

\begin{document}

\title{Online Harassment in Majority Contexts: Examining Harms and Remedies across Countries}

\author{Sarita Schoenebeck}
\affiliation{%
  \institution{University of Michigan}
  \country{USA}
}

\author{Amna Batool}
\affiliation{%
  \institution{University of Michigan}
  \country{USA}
}
\author{Giang Do}
\affiliation{%
  \institution{University of Michigan}
  \country{USA}
}
\author{Sylvia Darling}
\affiliation{%
  \institution{University of Michigan}
  \country{USA}
}
\author{Gabriel Grill}
\affiliation{%
  \institution{University of Michigan}
  \country{USA}
}
\author{Daricia Wilkinson}
\affiliation{%
  \institution{Clemson University}
  \country{USA}
}
\author{Mehtab Khan}
\affiliation{%
  \institution{Yale University}
  \country{USA}
}
\author{Kentaro Toyama}
\affiliation{%
  \institution{University of Michigan}
  \country{USA}
}
\author{Louise Ashwell}
\affiliation{%
  \institution{University of Michigan}
  \country{USA}
}

\renewcommand{\shortauthors}{Schoenebeck et al}
\begin{abstract}

Online harassment is a global problem. This article examines perceptions of harm and preferences for remedies associated with online harassment with nearly 4000 participants in 14 countries around the world. The countries in this work reflect a range of identities and values, with a focus on those outside of North American and European contexts. Results show that perceptions of harm are higher among participants from all countries studied compared to the United States. Non-consensual sharing of sexual photos is consistently rated as harmful in all countries, while insults and rumors are perceived as more harmful in non-U.S. countries, especially harm to family reputation. Lower trust in other people and lower trust in sense of safety in one's neighborhood correlate with increased perceptions of harm of online harassment. In terms of remedies, participants in most countries prefer monetary compensation, apologies, and publicly revealing offender's identities compared to the U.S. Social media platform design and policy must consider regional values and norms, which may depart from U.S. centric-approaches. 


\end{abstract}



\maketitle

\section{Introduction}
Online harassment is pervasive in regions around the world. Users post hate speech that demeans and degrades people based on their gender, race, sexual identity, or position in society \cite{blackwell2017classification,lenhart2016online}; users post insults and spread rumors, disproportionately harming those with fewer resources in society to cope with or respond to the attacks \cite{marwick2021morally, lenhart2016online, ybarra2008risky}; and users share private, sensitive content, like home addresses or sexual images, without the consent of those whose information is being shared \cite{goldberg2019nobody}. These behaviors introduce multiple types of harm with varied levels of severity, ranging from minor nuisances to psychological harm to economic precarity to life threats \cite{jiang2021understanding, schoenebeck2020reimagining, sambasivan2019they}. \textcolor{black}{Gaining a global understanding of online harassment} is important for designing online experiences that meet the needs of diverse, varied global experiences. 

Social media platforms have struggled to govern online harassment, relying on human and algorithmic moderation systems that cannot easily adjudicate content that is as varied as the human population that creates it \cite{goldman2021content, roberts2019behind}. Platforms maintain community guidelines that dictate what type of content is allowed or not allowed and then use the combination of human and automated pipelines to identify and address violations \cite{roberts2019behind, gillespie2018custodians}. However, identifying and categorizing what type of content is harmful or not is difficult for both humans and algorithms to do effectively and consistently. These challenges are magnified in multilingual environments where people may be trying to assess content in different languages or cultural contexts than they are familiar with, while algorithms are inadequately developed to work across these languages and contexts \cite{york2021silicon, gupta2022adima}. 

\textcolor{black}{Investigations} of harms associated with online harassment have been given disproportionate attention in U.S. contexts. Most prominent technology companies are centered in the U.S., employing U.S. workers in executive positions and centering U.S. laws, norms, corporations, and people \cite{york2021silicon, wef2022ceo}. Scholars have called attention to this problem, pointing out how experiences differ for people and communities globally (e.g. \cite{york2021silicon, sambasivan2019they, sultana2021unmochon}). For example, a study of 199 South Asian women shows that they refrain from reporting abuse because platforms rarely have the contextual knowledge to understand local experiences \cite{sambasivan2019they}. Across countries, social media users have expressed distrust in platforms' ability to govern behavior effectively, especially systems that are vague, complicated, and U.S.- and European-centric \cite{crawford2016flag, sambasivan2019they, blackwell2017classification}. 

Governing social media across the majority of the world requires understanding how to design platforms with policies and values that are aligned with the communities who use them. Towards that goal, this article examines perceptions of harm and preferences for remedies associated with online harassment via a survey conducted in 14 countries\footnote{\textcolor{black}{Data was collected from 13 countries plus a collection of Caribbean countries. We use the term "country" throughout for readability.}} around the world, selected for their diversity in location, culture, and economies. Results from this study shed light on similarities and differences in attitudes about harms and remedies in countries around the world. This work also demonstrates the complexities of measuring and making sense of these differences, which cannot be explained by a single factor and should not be assumed to be stable over time. This article advances scholarship on online harassment in majority contexts, and seeks to expand understandings about how to design platforms that meet the needs of the communities that use them. 

\section{Impacts of Online Harassment}
Online harassment is an umbrella term that encompasses myriad types of online behaviors including insults, hate speech, slurs, threats, doxxing, and non-consensual image sharing, among others. A rich body of literature has described characteristics of online harassment including what it is, who experiences it, and how platforms try to address it (e.g. \cite{schoenebeck2020reimagining, jhaver2019did, matias2019preventing, douek2020governing, chandrasekharan2017bag, thomas2021sok}). Microsoft's Digital Civility surveys and Google's state of abuse, hate, and harassment surveys indicate how harassment is experienced globally \cite{thomas2021sok, msft2022dci}. Harassment can be especially severe when it is networked and coordinated, where groups of people threaten one or many other people's safety and wellbeing \cite{marwick2021morally}. Other types of harassment are especially pernicious in-the-moment, such as reporting ``crimes'' so that law enforcement agencies investigate a home \cite{bernstein2016investigating} or sharing a person's home address online with the intent of encouraging mobs of people to threaten that person at their home. Across types of harm, marginalized groups experienced disproportionate harm associated with harassment online, including racial minorities, religious minorities, caste minorities, sexual and gender minorities, and people who have been incarcerated \cite{walker2017systematic, pewonline, powell2014blurred,maddocks2018non, englander2015coerced, poole2015fighting}. 

Sometimes users post malicious content intended to bypass community guidelines which are difficult to algorithmically detect \cite{dinakar2012common,vitak2017identifying}. This makes it relatively easy to deceive automatic detection models by subtly modifying an otherwise highly toxic phrase so that the detection model assigns it a significantly lower toxicity score \cite{hosseini2017deceiving}. In addition, due to limited training on non-normative behavior, these automatic detection and classification tools can exacerbate existing structural inequities \cite{hosseini2017deceiving}. For instance, Facebook’s removal of a photograph of two men kissing after flagging it as ``graphic sexual content'' highlighted the lack of inclusivity of non-dominant behavior in their automatic detection tools \cite{fbcencershipprob}. This valorization of certain viewpoints highlights that power resides among those who create these labels by embedding their own values and worldviews (mostly U.S.-centric) to classify particular behaviors as appropriate or inappropriate \cite{blackwell2017classification, hosseini2017deceiving}.

The effects of harassment vary by experience and individuals but might include anxiety, stress, fear, humiliation, self-blame, anger, and illness. There is not yet a standard framework for measuring harms associated with online harassment, which can include physical harm, sexual harm, psychological harm, financial harm, reproductive harm, and relational harm \cite{unwomen}. These can manifest in myriad ways: online harassment can cause changes to technology use or privacy behaviors, increased safety and privacy concerns, and disruptions of work, sleep, and personal responsibilities \cite{pittaro2007cyber,griffiths2002occupational,duggan2017online}. Other consequences can include public shame and humiliation, an inability to find new romantic partners, mental health effects such as depression and anxiety, job loss or problems securing new employment, offline harassment and stalking, numerous mental health issues, such as post-traumatic stress disorder (PTSD), depression, anxiety, self-blame, self-harm, trust issues, low self-esteem, confidence, and loss of control \cite{walker2017systematic, powell2014blurred, bates2017revenge,eckert2020doxxing, barrense2020non, ryan2018european, pampati2020having}. These effects can be experienced for long periods of time due in part to the persistence and searchability of content \cite{goldberg2019nobody}. Targets often choose to temporarily or permanently abstain from social media sites, despite the resulting isolation from information resources and support networks \cite{goldberg2019nobody, lenhart2016online}. 

Microsoft's Digital Civility Index, a yearly survey of participants in over 20 countries, indicates that men are more confident than women in managing online risks \cite{msft2022dci}. Sexual images of women and girls are disproportionately created, sent, and redistributed without consent which can severely impact women's lives \cite{burkett2015sex, dobson2016sext, bates2017revenge,eckert2020doxxing, barrense2020non}. In a study of unsolicited \textcolor{black}{nude} images and its affect on user engagement \cite{hayes2018unsolicited, shaw2016bitch}, victims reported being bombarded with unwelcomed explicit imagery and faced further insults when they attempted to reduce interaction. A survey by Maple et al. with 353 participants from the United Kingdom (68\% of respondents were women) listed damage to their reputation as the primary fear of victims of cyberharassment \cite{maple2011cyberstalking}. 

The consequences of gendered and reputational harm can be devastating. In South Korea, celebrities Hara Goo and Sulli (Jin-ri Choi) died by suicide, which many attributed to the large-scale cyberbullying, sexual harassment, and gender violence they experienced online \cite{goo2019}. A social media Pakistani celebrity was murdered by her brother, who perceived her social media presence as a blemish on the family's honor \cite{QandeelBaloch}. Two girls and their mother were allegedly gunned down by a stepson and his friends over the non-consensual filming and sharing of a video of the girls enjoying rain among family \cite{twogirls}.  
Many of these harms are ignited and fueled by victim-blaming, where society places the responsibility solely on women and other marginalized groups to avoid being assaulted \cite{walker2017systematic, powell2014blurred, chisala2016gender}. This blaming is also perpetuated digitally; for instance, a review of qualitative studies on non-consensual sharing highlighted that women are perceived as responsible if their images are shared because they voluntarily posed for and sent these images in the first place \cite{walker2017systematic}.

\section{Challenges in Governing Online Harassment}
Most social media sites have reporting systems aimed at flagging inappropriate content or behavior online \cite{crawford2016flag}. Though platform policies do not explicitly define what constitutes online harassment \cite{pater2016characterizations}, platforms have highlighted several activities and behaviors in their community guidelines including abuse, bullying, defaming, impersonation, stalking, and threats \cite{pater2016characterizations, jiang2020characterizing}. Content that is reported goes into a processing pipeline where human workers evaluate the content and determine whether it violates community guidelines or not \cite{roberts2019behind}. If it does, they may take it down and sanction the user who posted it, with sanctions ranging in severity from warnings to suspensions to permanent bans \cite{schoenebeck2021drawing, goldman2021content}. Platforms use machine learning to automatically classify and filter out malicious content, abusive language, and offensive behaviors \cite{chandrasekharan2017bag,wulczyn2017ex,yin2009detection}. 
These range from adding contextual and semantic features in detection tools to generating computational models using preexisting data from online communities to using these machine learning models to assign ``toxicity scores'' \cite{wulczyn2017ex, chandrasekharan2017bag}. Though harassment detection approaches have improved dramatically, fundamental limitations remain \cite{blackwell2017classification}, including false positives and true negatives, where content is taken down that should have stayed up and vice versa \cite{haimson2021disproportionate, schoenebeck2020reimagining}. 

Many of these problems are deeply embedded in algorithmic systems, which can reinforce Western tropes, such as associating the word "Muslim" with terrorists \cite{abid2021persistent}. Algorithms to detect problematic content also perform substantially worse in non-English languages, perpetuating inequalities rather than remediating them \cite{debre2021facebook}. Dominant voices can overrule automatically detected flagged content through situated judgments \cite{crawford2016flag}. For instance, a widely distributed video of Neda, an Iranian woman caught up in street protests and shot by military police in 2009, was heavily flagged as violating YouTube's community guidelines for graphic violence, but YouTube justified leaving it up because the video was newsworthy \cite{Neda}. 

Platform policies are written in complex terms that are inaccessible to many social media users, which makes it difficult for them to seek validation of their online harassment experiences \cite{fiesler2016reality}. Further, platform operators do not specify which prohibited activities are associated with which responses \cite{pater2016characterizations}. When combined with the punitive nature of sanctions, online governance systems may be confusing and ineffective at remediating user behavior, while overlooking the harms faced by victims of the behavior \cite{schoenebeck2021drawing}. One alternative that has been proposed more recently is a focus on rehabilitation and reparation in the form of apologies, restitution, mediation, or validation of experiences \cite{blackwell2017classification, schoenebeck2021drawing, xiao2022sensemaking}. Implementing responses to online harassment requires that users trust platforms' ability to select and implement that response \cite{wilkinson2022many}; however, public trust in technology companies has decreased in recent years, and there is also distrust of social media platforms' ability to effectively govern online behavior \cite{schoenebeck2021youth, americanstrust2020, blackwell2017classification, musgrave2022experiences}. 84\% of social media users in the U.S. believe that it is the platform's responsibility to protect them from social media harassment \cite{Americans}, yet Lenhart et al.'s survey suggests that only 27\% of victims reported harassing activities on these platforms \cite{onlineharassmentAmerica}. A different survey by Wolak et al. with 1631 victims of sextortion found that 79\% of victims did not report their situation to social media sites because they did not think it would be helpful to report \cite{wolak2016sextortion}. Their participants indicated that platform reporting might be helpful only when victims are connected to perpetrators exclusively online which might be addressable through in-app reporting \cite{wolak2016sextortion}. Sambasivan et al.'s study with 199 South Asian women revealed that participants refrain from reporting through platforms due to platforms' limited contextual understanding of victims' regional issues, which is further slowed by the platforms' requirements to fill out lengthy forms providing detailed contexts \cite{sambasivan2019they}. Musgrave et al. find that U.S. Black women and femmes do not report gendered and racist harassment because they do not believe reporting will help them \cite{musgrave2022experiences}.

Wolak et al. also found that only 16\% of victims of sextortion reported their incidents to the police \cite{wolak2016sextortion}. Many of those who reported to police described having a negative reporting experience, which deterred them from pursuing criminal charges against offenders \cite{wolak2016sextortion}. Such experiences include police arguing for the inadequacy of proof to file complaints, that sextortion is a non-offensive act, lack of jurisdiction to take actions, and being generally rude, insensitive, and mocking \cite{wolak2016sextortion}. Sambasivan et al. also reported that only a few of their nearly 200 participants reported abusive behaviors to police because they perceived law enforcement officers to have low technical literacy, to be likely to shame women, or to be abusers themseves. \cite{sambasivan2019they}. 
When abusers are persistent, even reporting typically does not address the ongoing harassment \cite{marwick2021morally, goldberg2019nobody}. 


 Sara Ahmed introduces the concept ``strategic inefficiency'' to explain how institutions slow down complaint procedures that can then deter complaints from constituents \cite{ahmed2021complaint}. The lack of formal reporting channels leads users to be largely self-reliant for mitigating and avoiding abuse. Techniques they use range from preventative strategies like limiting content, modifying privacy settings, self-censorship, using anonymous and gender-neutral identities, using humor, avoiding communication with others, ignoring abuse, confronting abusers, avoiding location sharing, deleting accounts, blocklists, changing contact information, changing passwords, using multiple emails accounts for different purposes, creating a new social media profile under a different name, blocking or unfriend someone and untagging themselves from photos \cite{onlineharassmentAmerica, wolak2016sextortion, vitak2017identifying, fox2017women, mayer2021now, such2017photo, corple2016beyond, dimond2013hollaback, vitis2017dick, jhaver2018online}.  Whether reporting to companies or \textcolor{black}{police}, these approaches all put the burden of addressing harassment on the victims. If we want to better govern online behavior globally, we need to better understand what harms users experience and how platforms and policies can systematically better support them after those harms.

\section{Study Design}
We conducted a cross-country online survey in 14 countries (13 countries plus multiple Caribbean countries). \textcolor{black}{We aimed for a minimum of 250 respondents in each country which considered our desire for age variance and gender representation among men and women but without the higher sample size needed for representative samples or subgroup analyses}. The survey focused on online harassment harms and remedies and included questions about demographics, personal values, societal issues, social media habits, and online harassment. This paper complements a prior paper from the same project that focused on gender \cite{im2022women}; this paper focuses on country level differences though it also engages with gender as part of the narrative. 

We iteratively designed the survey as a research team, discussing and revising questions over multiple months. When we had a stable draft of a survey, members of our research team translated surveys manually and compared those versions to translations via paid human translation services for robustness. We pilot tested translations with 2-4 people for each language and revised the survey further. \textcolor{black}{Our goal was to have similar wording across languages; though this resulted in some overlapping terms in the prompts (e.g. malicious), participants seemed to comprehend each prompt in our pilots.} We deployed the survey in a dominant local language for each country (see Table \ref{table:participant-demographics}). The survey contained four parts: harassment scenarios, harm measures, possible remedies, and demographics and values. Below, we describe each stage in detail:

\textit{Harassment scenarios}. We selected four online harassment scenarios to capture participants' perceptions about a range of harassment experiences but without making the survey too long which leads to participant fatigue. We selected the four harassment scenarios by reviewing prior scholarly literature, reports, and \textcolor{black}{news articles} and prioritizing diversity in types of harm and severity of harm. We prioritized harassment types that would be globally relevant and legible among participants and could be described succinctly. Participants were presented with one scenario along with the harm and remedy questions (described below), and completed this sequence four times for each harassment scenario. The harassment scenario prompt asked participants to "Imagine a person has:" and then presented each of the experiences below. 
\begin{itemize}
\item spread malicious rumors about you on social media
\item taken sexual photos of you without your permission and shared them on social media
\item insulted or disrespected you on social media
\item created fake accounts and sent you malicious comments through direct messages on social media
\end{itemize}

\textit{Harm measures}. We developed four measures of harm to ask about with each harassment scenario. We again prioritized types of harmful experiences that would be relevant to participants globally. Drawing on our literature review on harms in other disciplines (e.g. medicine) and more nascent discussions of technological harms (e.g. privacy harms \cite{citron2021privacy}), we chose to prioritize three prominent categories of harm used in scholarly literature and by the World Health Organization--psychological, physical, and sexual harm. We then added a fourth category---reputational harm---because harm to family reputation is a prominent concern in many cultures and these concerns may be exacerbated on social media. We prioritized question wording that could be translated and understood across languages. For example, our testing revealed that the concept of ``physical harm'' was confusing to participants when translated so we iterated on wording until we landed on personal safety. The final wording we used was:

\begin{itemize}
\item Would you be concerned for your psychological wellbeing?
\item Would you be concerned for your personal safety?
\item Would you be concerned for your family reputation?
\item Would you consider this sexual harassment against you?
\end{itemize}

Perceived harm options were presented on 5-point scales of ``Not at all concerned'' (1) to ``Extremely concerned'' (5) for the first three questions and ``Definitely not'' (1) to ``Definitely'' (5) for the last question. We chose these response stems to avoid Agree/Disagree options which may promote acquiescence bias \cite{saris2010comparing} and because these could be translated consistently across languages. 

\textit{Harassment remedies}. Current harassment remedies prioritize content removal and user bans after a policy violation. However, scholars are increasingly arguing that a wider range of remedies is needed for addressing widespread harms. Goldman proposes that expanded remedies can improve the efficacy of content moderation, promote free expression, promote competition among Internet services, and improve Internet services’ community-building functions \cite{goldman2021content}. Goldman's taxonomy of remedies is categorized by content regulation, account regulation, visibility reductions, monetary, and ``other.'' Schoenebeck et al. \cite{schoenebeck2020drawing} have also proposed that expanding remedies can create more appropriate and contextualized justice systems online. They see content removal and user bans as a form of criminal legal moderation, where harmful behavior is removed from the community, and propose adding complementary justice frameworks. For example, restorative justice suggest alternative remedies like apologies, education, or mediation. Building on this work, we developed a set of proposed remedies and for each harassment scenario, we asked participants, ``How desirable would you find the following responses?'' with response options on a 5-point scale of ``Not at all desirable for me (1)'' to ``Extremely desirable for me (5).'' The seven remedies we displayed were chosen to reflect a diversity of types of remedies while keeping the total number relatively low to reduce participant fatigue. We also asked one free response question ``What do you think should be done to address the problem of harassment on social media?''

\begin{itemize}
\item removing the content from the site.
\item labeling the content as a violation of the site’s rules.
\item banning the person from the site.
\item paying you money.
\item requiring a public apology from the person.
\item revealing the person’s real name and photograph publicly on the site.
\item by giving a negative rating to the person.
\end{itemize}

\textit{Demographics}. The final section contained social media use, values, and demographic questions. The values and demographic questions were derived from the World Values Survey (WVS) \cite{inglehart2014world}, a long-standing cross-country survey of values. This paper focuses on six measures from the WVS. 

\begin{itemize}
\item Generally speaking, would you say that most people can be trusted or that you need to be very careful in dealing with people? 
\item How much confidence do you have in police?
\item How much confidence do you have in the courts?
\item How secure do you feel these days in your neighborhood?
\item What is your gender? 
\item Have you had any children?
\end{itemize}

The response options ranged from ``None at all'' (1) to ``A great deal'' (4) for police and courts and from Not at all secure (1) to Very secure (4) for neighborhood. We omitted the police and courts questions in Saudi Arabia. For trust, options were ``Most people can be trusted'' (1) and ``Need to be very careful'' (2). For gender, the response options were ``Male'', ``Female'', ``Prefer not to disclose'', and ``Prefer to self-describe.'' We chose not to include non-binary or transgender questions because participants in some countries cannot safely answer those questions, though participants could choose to write them in.


We recruited participants from 14 countries (see Table \ref{table:participant-demographics}): 13 countries plus the Caribbean countries (Antigua and Barbuda, Barbados, Dominica, Grenada, Jamaica, Monserrat, St. Kitts and Nevis, St. Lucia, and St. Vincent). \textcolor{black}{We decided to analyze the Caribbean countries together because of the small sample sizes and their relative similarities, while recognizing that each country has its own economics, culture and politics.} This study was exempted from review by our institution’s Institutional Review Board. Participants completed a consent form in the language of the survey. Participants were recruited via the survey company Cint in most countries, Prolific in the U.S., and manually via the research team in the Caribbean countries and Mongolia. Participants were compensated based on exchange rates and pilot tests of time taken in each country. 

\begin{table}
\caption{Participant demographics}
\label{table:participant-demographics}
\begin{tabular}{ l c c  }
\toprule Country & Language & Num Participants\\  \midrule
Austria & German & 251 \\
Cameroon & English & 263   \\
Caribbean & English & 254 \\
China & Mandarin & 283 \\
Colombia & Spanish (Colombian) & 296 \\
India & Hindi/English & 277 \\
South Korea & Korean & 252 \\
Malaysia & Malay & 298 \\
Mexico & Spanish (Mexican) & 306 \\
Mongolia & Mongolian & 367 \\
Pakistan & Urdu & 302 \\
Russia & Russian & 282 \\
Saudi Arabia & Arabic & 258 \\
USA & English & 304 \\
\textbf{Total} & & \textbf{3993}
\end{tabular}
\end{table}

\subsection{Participant Demographics} The gender ratio between men and women participants was similar across countries \textcolor{black}{ranging from 50\% women and 50\% men in China to 43\% women and 57\% men in India)} except for Caribbean countries \textcolor{black}{which was women: 69\%, men: 27\% and Mongolia which was women: 59\%, men: 41\%} (see details about gender in \cite{im2022women}). The median age was typically in the 30s; Mongolia was lowest at 21 while South Korea and United States were 41.5 and 44, respectively. Participants skew young but roughly reflect each country's population, e.g. Mongolia’s median age is 28.2 years while South Korea and U.S medians are 43.7 and 38.3, respectively, according to United Nations estimates \cite{united20192019}. Participants’ self-reported income also varied across countries, with participants in Austria reporting higher incomes and participants in Caribbean countries reporting lower incomes. More than half of the participants had education equivalent to a Bachelor degree for eight countries (Cameroon, China, Colombia, India, Malaysia, Russia, Saudi Arabia, United States); the other countries did not. Participants placed their political views as more ``left'' than ``right.'' 

\subsection{Data analysis} We discarded low-quality responses based on duration (completed too quickly) and data quality (too many skipped questions). Table \ref{table:participant-demographics} shows the final number of participants per country after data cleaning. For the qualitative analysis, we separately discarded responses that were low quality (empty fields, meaningless text); the number of participants was slightly higher overall (N=4127) since some participants completed that section but did not finish the subsequent quantitative portions of the survey. 

We analyzed data using R software. We used group means to describe perceived harms and preferred remedies. Levene's tests to measure variance were significant for both harm and remedy analyses indicating that homogeneity of variance assumption is violated. Thus, we used Welch one-way tests for nonparametric data and posthoc pairwise t-tests which we deemed appropriate given our sufficiently large sample size \cite{fagerland2012t}. We used the Benjamini–Hochberg (BH) test to correct for multiple comparisons \cite{bretz2016multiple}. We also ran linear regressions with harassment - harm and harassment - remedy pairings as the dependent variables and demographics and country as the independent variables (4 harassment scenarios x 4 harm types = 16 harm models; 4 harassment scenarios x 7 remedies = 28 remedy models). We used adjusted R-squared to identify demographic variables that were more likely to explain model variance. Welch test and posthoc tests for harm (16 harassment-harm pairings) and remedy (28 harassment - remedy pairings) comparisons are available in the Appendix. Regression outputs and confidence intervals for demographic predictors are also available in the Appendix. 

We analyzed the qualitative data to the free responses question using an iterative, inductive process. Our approach was to familiarize ourselves with the data, develop a codebook, iteratively refine the codebook, code the data, then revisit the data to make sense of themes from the coding process. To do this, four members of the research team first read through a sample of responses across countries and then co-developed a draft codebook. Three members of the team then coded a sample of responses and calculated interrater reliability (IRR) for each code using Cohen's Kappa. Across the 26 codes tested, Kappa values ranged from -0.1 to 1 with a median of .35. We used the IRR values as well as our manual review of differences to refine the codebook. We removed codes that coders did not interpret consistently, generally those with agreement below about .4 and those that were low prevalence in the data. We revised remaining codes, especially those that had lower agreement, and discussed them again. The final codebook contained 21 codes (see Appendix) that focused on moderation practices, user responsibility, government involvement, and other areas of interest. 

\subsection{Limitations and Mitigation}
Cross-country surveys are known to have a range of challenges that are difficult to overcome completely, but they remain useful, even indispensable, if designed and interpreted thoughtfully and cautiously~\cite{kaminska2017survey,kish1994multipopulation,smith2011opportunities}. 

In our case, the key issues have to do with language, sampling methodologies, and response biases that might have differed across our participants. Language differences were addressed as described above, through a process of careful translation, validation through back-translation, and survey piloting, but topics like non-consensual image sharing are inevitably shaped by the language they are discussed in and there may be differences in interpretation we did not capture. Sampling methodologies within countries were as consistent as we could make them, but a number of known differences should be mentioned: First, we used three different mechanisms for recruiting -- a market research firm (Cint) for 11 countries; a research survey firm (Prolific) for the United States; and our own outreach for the Caribbean and Mongolia. These mechanisms differ in the size of their pool of participants, as well as their baseline ability to draw a representative sample. Some differences were built-in to the recruitment process, for example, we requested a diverse age range of participants explicitly with Cint and Prolific which should have yielded more older adults. In contrast, our researcher recruitment method for Caribbean and Mongolia simply sought a range of participants through word of mouth, but did not specifically recruit or screen for older adults. Second, while we sought representative samples of the national/regional population in all cases, we know that we came up short. For example, while online surveys are increasingly able to achieve good representation in better-educated countries with high internet penetration, 
they are known to be skewed toward affluent groups in lower-income, less-connected contexts~\cite{mohorko2013internet,tijdens2016web}. \textcolor{black}{Oversampling from groups who are active online may be more tolerable for a study of online harassment, but it still overlooks important experiences from those who may be online but less likely to participate in a survey.} Third, differences in local culture and current events are known to cause a range of response biases across countries. Subjective questions about perception of harm, for example, might depend on a country's average stoicism; questions about ``trust in courts'' might be affected by the temporary effects of high-profile scandals. The issues above are common to cross-country survey research, and our mitigation strategies are consistent with the survey methodology literature ~\cite{kaminska2017survey,kish1994multipopulation}. 

To provide some assurance of our data's validity, we benchmarked against the World Values Survey, on which some of our demographic and social-issues questions were based. We compared responses from our participants to responses from the WVS for countries that had WVS data (China, Colombia, partial India, South Korea, Malaysia, Mexico, Pakistan, Russia, United States). We used the more recent Wave 7 (2017-20) where data was available, with Wave 6 (2010-14) as a back-up. We expected that our responses should correlate somewhat with WVS, even though there were substantial differences, such as that our sample was recruited via online panels with questions optimized for mobile devices whereas the WVS sample was recruited door-to-door with oral question and answer choices. Sample means for our data and the WVS for similar questions are presented in plots in the Appendix. In countries where corresponding data is available, we find that the means in our data about trust -- in police, or in courts -- align with WVS results. We also find the anticipated biases with respect to online surveys and socio-economic status. In particular, our participants reported better health and more appreciation for gender equality than WVS participants. 

Still, because of the above, we present our results with some caution, especially for between-country comparisons; specific pair-wise comparisons between countries should be considered with substantial caution. We include specific comparisons primarily in the Appendix for transparency; we focus on patterns in the Results which we expect to be more reliable, especially patterns within countries and holistic trends across the entire dataset. In the following sections, we strive to be explicit about how our findings can be interpreted.

\section{Results}
Results are organized into two sections: perceptions of harm associated with online harassment and preferences for remedies associated with online harassment. Each section follows the same structure: first we look at which harassment types are perceived as most harmful and which remedy types are most preferred, respectively, then we examine demographic predictors of perceptions of harm and preferences for remedies, respectively. 

\subsection{Perceptions of Harm Associated with Online Harassment}

\begin{figure}
    \centering
    \includegraphics[width=.7\linewidth]{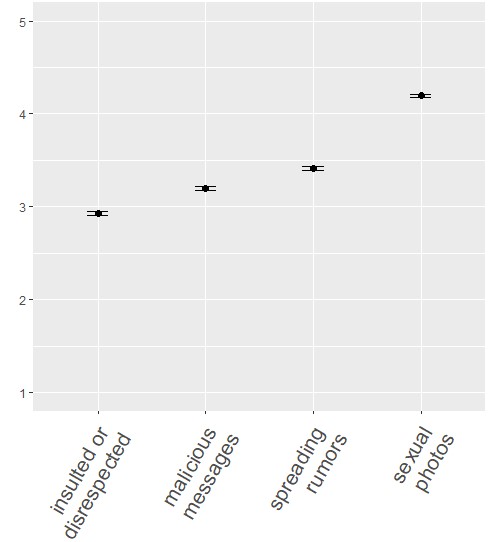}
    \caption{Perceptions of harm by harassment type}
    \label{fig:harm_plot}   
    \Description{Plot with error bars from lowest harm to highest: insulted or disrespected, malicious messages, spreading rumors, sexual photos.}
\end{figure}

First, we differentate between the four types of harassment. Figure \ref{fig:harm_plot} shows perceptions of overall harm by harassment type. One-way Welch tests showed that means of perceptions of harm were significantly different, $F$(3, 35313) = 3186.4, $p$ < 0.001, with sexual photos being the highest in harm (M=4.20, SD=1.15), followed by spreading rumors (M=3.42; SD=1.30), malicious messages (M=3.20; SD=1.35), and insults or disrespect (M=2.93; SD=1.36) (see Figure \ref{fig:harm_plot}). Plots and posthoc tests for comparisons by type of harassment by country are available in the Appendix.

To display an overall measure of perceived harms associated with online harassment by country, we aggregated each of the four harm measures together -- sexual harassment, psychological harm, physical safety, and family reputation -- for a combined measure of overall harm.

\begin{figure}
    \centering
    \includegraphics[width=\linewidth]{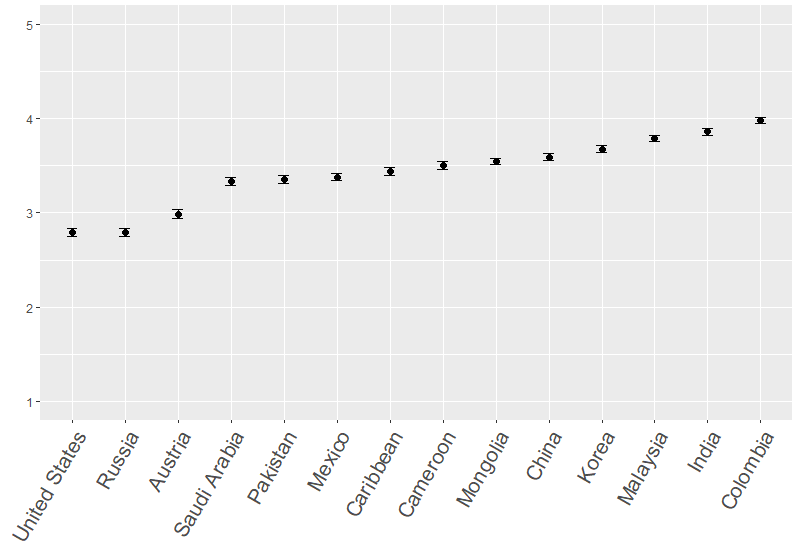}
    \caption{Perceptions of harm by country}
    \label{fig:harm_region}
    \Description{Plot with error bars from lowest harm to highest: United States, Russia, Austria, Saudi Arabia, Pakistan, Mexico, Caribbean, Cameroon, Monoglia, China, Korea, Malaysia, India, Colombia.}
\end{figure}

Results suggest that participants in Colombia, India, and Malaysia rated perceived harm highest, on average, while participants in the United States, Russia, and Austria perceived it the lowest. Means are presented here and shown visually in Figure \ref{fig:harm_region}: Colombia (M=3.98; SD=1.18); India (M=3.86; SD=1.31); Malaysia (M=3.79; SD=1.21); Korea (M=3.67; SD=1.22); China (M=3.59; SD=1.19); Mongolia (M=3.55; SD=1.29); Cameroon (M=3.50; SD=1.38); Caribbean (M=3.44; SD=1.43); Mexico (M=3.38; SD=1.38); Pakistan (M=3.36; SD=1.36); Saudi Arabia (M=3.34; SD=1.35); Austria (M=2.99; SD=1.42); Russia (M=2.80; SD=1.43); United States (M=2.79; 1.45).

Most of the ratings by country were statistically significant from each other (one-way Welch tests and posthoc tests are reported in the Appendix), though we remind readers that these differences should be interpreted with caution. In general, the wealthier countries per capita perceive lower harm, but beyond that the key takeaway is that there is substantial variance which is unlikely to be explained by one or even a few differences across any of those countries.

\subsubsection{Predictors of Perceptions of Harm}
\begin{figure}[ht]
    \centering
    \includegraphics[width=1.05\linewidth]{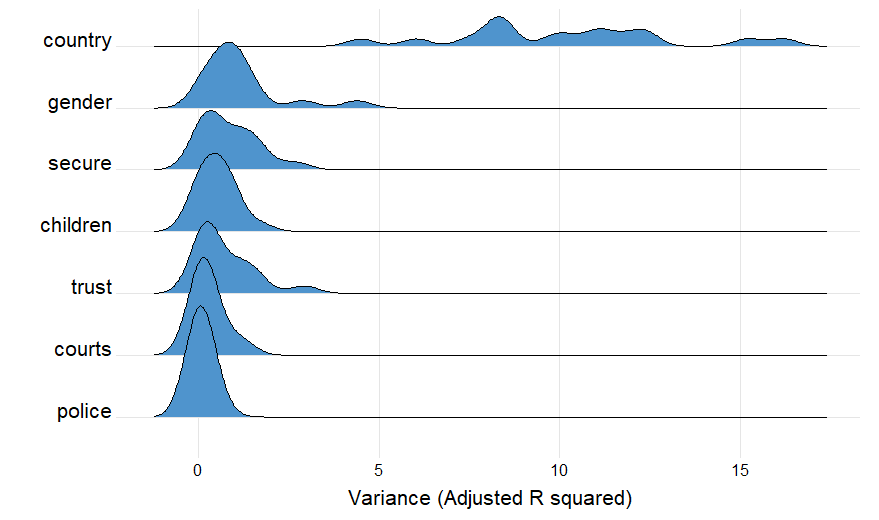} 
    \caption{Adjusted R squared of demographic variables for predicting harm across 16 harassment scenario x harm types.}
    \label{fig:harm_ridgeline}   
    \Description{Ridgeline plot (looks like a wave with one or a few peaks) showing which variables predict harm from highest to lowest: country, gender, secure, children, trust, courts, police.}
\end{figure}

Here we hone in more granular differences across harassment types and harm types and how they vary by country and other demographic data. 
Note that responses from Saudi Arabia participants are excluded from regressions because they did not complete questions about confidence in courts or police. The distribution of R-squared values for the 16 harassment - harm pairings is shown in Figure \ref{fig:harm_ridgeline} (ranging from close to 0 to 18\% variance). Country was the most predictive of perception of harm though with variance across harassment and harm pairings as indicated by the multiple peaks in Figure \ref{fig:harm_region}. Gender was next most predictive, followed by security in neighborhood, number of children, trust of people, trust in courts, and trust in police. 

We also ran exploratory factor analyses to look for underlying constructs across measured variables. When all variables we measured were in the analysis, perceptions of harm and preferred remedies loaded into constructs, as expected, but demographic and value variables did not. Analyses with only the demographic and values variables suggest some trends but they were not substantial predictors of variance (e.g. trust and courts loaded together; marriage and age inversely loaded together). We show some factor analyses results in the Appendix but do not focus on them further. 


\begin{figure*}[ht]
  \centering
  \includegraphics[width=0.98\linewidth]{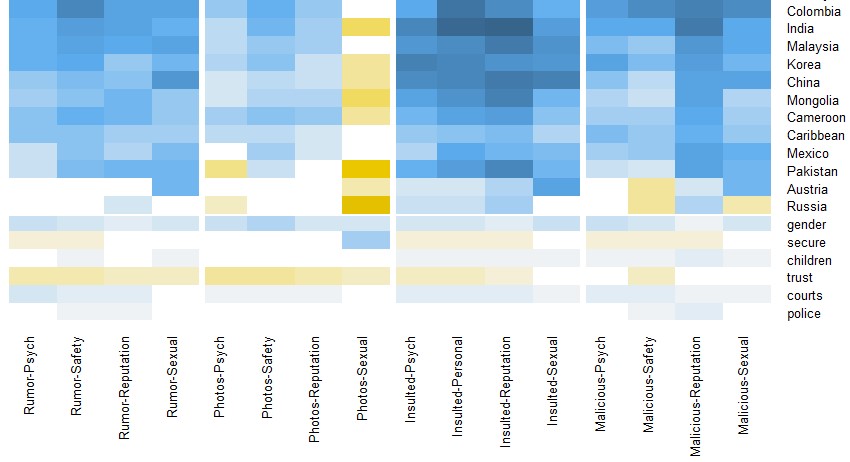}
  \caption{Heatmap of regression coefficients of harassment types and harm pairings by country and demographics. Darker blue is positive coefficient (i.e. higher harm); darker gold is negative coefficient.}
  \Description{Heatmap showing darker blue shades for countries, especially for the insult scenario. One exception is the photos scenario and sexual harassment which has gold (i.e. negative) coefficients.}
  \label{figure:harm_heatmap}
\end{figure*}

We ran regression analyses for the 16 harassment type - harm pairings using country, gender, security in neighborhood, number of children, trust in other people, trust in courts, and trust in police as independent variables. We used the U.S. as the reference choice for country and men as the reference for gender. Complete results with confidence intervals are available in the Appendix. To communicate patterns across models, we present a heatmap (see Figure \ref{figure:harm_heatmap}) of regression coefficients with harassment type - harm pairings on the x-axis and the predictors in Figure \ref{fig:harm_ridgeline} on the y-axis. We also plotted participant responses to the courts, police, security, and trust questions with WVS ratings to benchmark that our participants' attitudes reflect those of a broader population; those plots are in the Appendix. 

Results of the six predictors in the regression models are summarized here:

\textit{Country}: Participants in most countries perceive higher harm for most pairings than the U.S., with the exception of \textcolor{black}{the sexual photos and sexual harm pairing} where some countries perceive lower harm than the U.S. 

\textit{Gender}: Women perceive greater harm than men for all 16 harassment - harm pairings. 

\textit{Secure}: Participants who were more likely to give low ratings to the question ``How secure do you feel these days in your neighborhood?'' were more likely to perceive higher harm associated with online harassment for 8 of the harassment - harm pairings; however, security in neighborhood is negatively correlated with ratings for the sexual photos - sexual harassment pairing.

\textit{Children}: Having more children is a predictor of greater perceptions of harm for 9 of the 16 pairings, except for the insulted or disrespected - sexual harassment pairing which is negatively correlated. 

\textit{Trust}: Participants who were more likely to be low in trust of other people were more likely to perceive higher harm associated with online harassment for 11 of the 16 harassment - harm pairings. The relationship was stronger for the sexual photos and spreading rumors scenarios, whereas there were no relationships for the malicious harassment scenario.

\textit{Courts}: Higher trust in courts is correlated with increases in perceptions of harm for 14 of the 16 pairings. The two exceptions are spreading rumors - sexual harassment and sexual photos - sexual harassment pairings. 

\textit{Police}: Trust in police is correlated with increases in perceptions of harm for 4 of the 16 pairings. 

We return to these results in the Discussion. 

\subsection{Preferences for Remedies Associated with Online Harassment}

The prior section presented perceptions of harm; this section presents preferences for remedies. Specifically we report respondents' perceived desirability of the remedies to address harassment - related harms. 

\begin{figure}
    \centering
    \includegraphics[width=.9\linewidth]{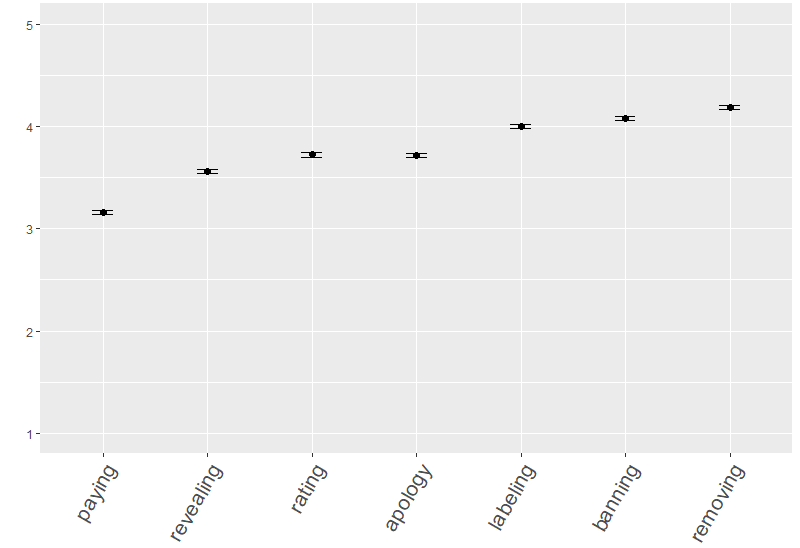}
    \caption{Preferences for remedies by
    \textcolor{black}{remedy} type}
    \label{fig:remedy_plot}      
    \Description{Plot with error bars from lowest remedy preference to highest: paying, revealing, rating, apology, labeling, banning, removing.}
\end{figure}

First, we differentiate between the remedies themselves. One-way Welch tests showed that means of preferences for remedies were significantly different, $F$(6, 49593) = 1130.9, $p$ < 2.2e-16 (see Figure \ref{fig:remedy_plot}). Removing content and banning offenders are rated highest, followed by labeling, then apologies and rating. Revealing identities and payment are \textcolor{black}{rated} lowest. Posthoc comparisons showed that all pairings were significantly different from each other except for apology and rating: removing (M=4.18; SD=1.12); banning (M=4.07; SD=1.17); labeling (M=4.00; SD=1.19); apology (M=3.72; SD=1.34); rating (M=3.72; SD=1.32); revealing (M=3.56; SD=1.39); paying (M=3.16; SD=1.46). 

\begin{figure}
    \centering
    \includegraphics[width=\linewidth]{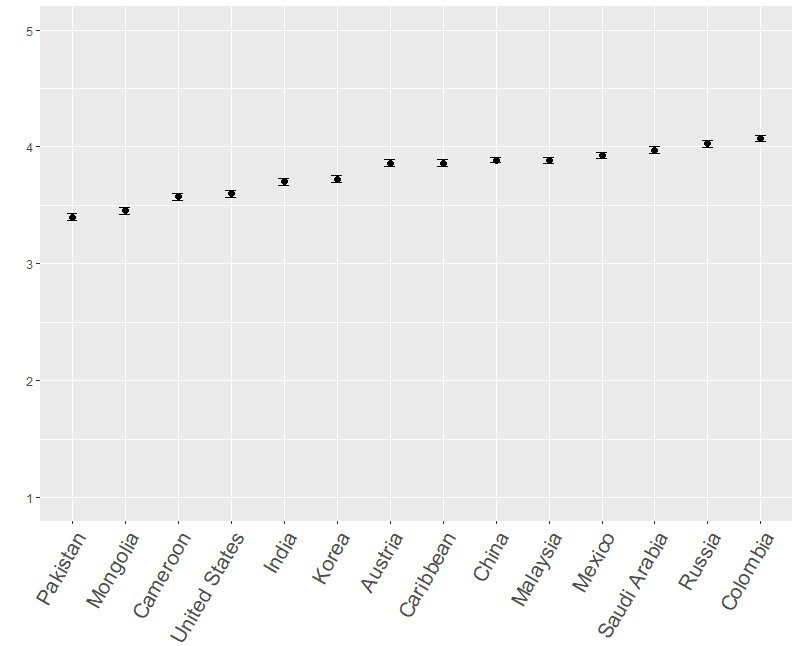}
        \caption{Preferences for remedies by country}
        \label{fig:remedy_region}   
        \Description{Plot with error bars from lowest remedy preference to highest: Pakistan, Mongolia, Cameroon, United States, India, Korea, Austria, Caribbean, China, Malaysia, Mexico, Saudi Arabia, Russia, Colombia.}
\end{figure}

To display overall preferences for remedies associated with online harassment by country, we aggregate the seven remedy types together for a combined measure of overall remedies. 
Results show that Colombia, Russia, and Saudi Arabia were highest overall in support for remedies while Pakistan, Mongolia, and Cameroon were lowest. Means are again presented here and shown visually in Figure \ref{fig:remedy_region}: Colombia (M=4.07; SD=1.13); Russia  (M=4.03; SD= 1.25); Saudi Arabia (M=3.97; SD=1.27); Mexico (M=3.93; SD=1.25); Malaysia (M=3.89; SD=1.20); China (M=3.89; SD=1.07); Caribbean (M=3.86; SD=1.38); Austria (M=3.86; SD=1.36); Korea (M=3.72; SD=1.29); India (M=3.70; SD=1.39); United States (M=3.60; SD=1.50); Cameroon (M=3.57; SD=1.40); Mongolia (M=3.45; SD=1.40); Pakistan (M=3.40; SD=1.41). 
As with harms, most of the ratings by country were statistically significant from each other (one-way Welch tests and posthoc tests are reported in the Appendix), though we again caution that the differences should be treated with caution. Regression outputs and confidence intervals for demographic predictors are also available in the Appendix.


\subsubsection{Predictors of Preferences for Remedies} 

\begin{figure}[ht]
    \centering
    \includegraphics[width=1.05\linewidth]{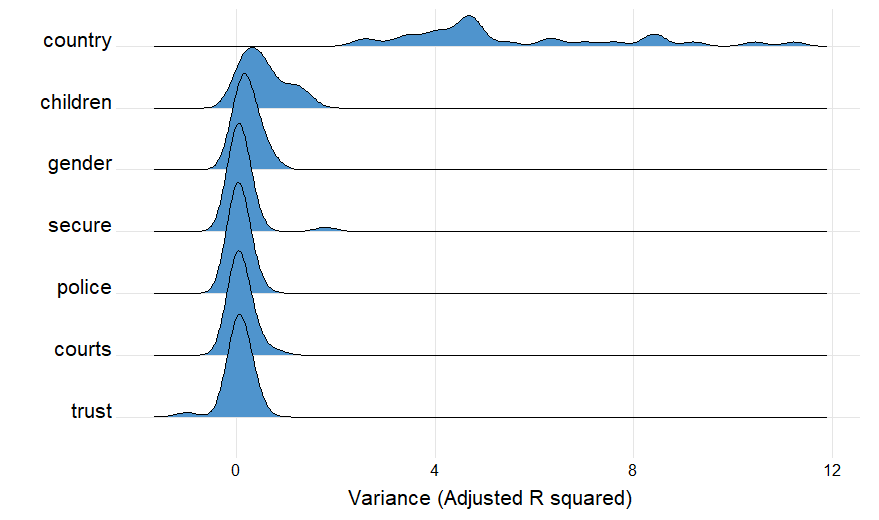} 
    \caption{Adjusted R squared of demographic variables for remedy preferences across 28 harassment scenario x remedy types.}
    \Description{Ridgeline plot (looks like a wave with one or a few peaks) showing which variables predict remedy preferences from highest to lowest: country, children, gender, secure, police, courts, trust.}
    \label{fig:remedy_ridgeline}    
\end{figure}

We plotted R-squared values with the same variables used in the harm regression models (see Figure \ref{fig:remedy_ridgeline}). Results are broadly similar to the harm ridgeline plot, though there is less overall variance explained in the remedy plots (0-15\%). Country is most predictive of preference for remedy, followed by number of children, gender, security in neighborhood, trust in police, trust in courts, and trust in other people. We ran regression analyses for the 28 harassment type - remedy pairings (4 harassment types and 7 remedies). 

\begin{figure*}[ht]
  \centering
  \includegraphics[width=0.98\linewidth]{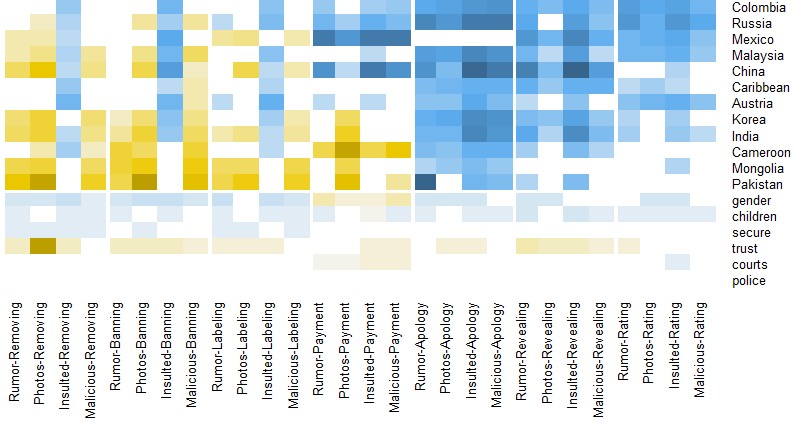}
  \caption{Heatmap of regression coefficients of harassment types and remedy pairings by country and demographics. Darker blue is positive coefficient (i.e. higher preference for remedy); darker gold is negative coefficient.}
  \Description{Heatmap showing preferences for remedies. The heatmap has small squares whose color reflects coefficient value for that variable. An overall visual pattern is that apology, revealing, and rating tend to be dark blue meaning that most countries prefer them to the U.S. The other remedies are gold or more mixed.}
  \label{fig:remedy_heatmap}
\end{figure*}

We visually show model results in a heatmap (see Figure \ref{fig:remedy_heatmap}). Results are summarized here:

\textit{Country:} Most countries tend to prefer payment, apologies, revealing users, and rating users, but are less favorable towards removing content, labeling content, or banning users compared to the U.S. These patterns are observed for three of the four harassment types, with the exception of insults or disrespect where countries tend to prefer all remedies compared to the U.S. 

\textit{Gender:} Women tend to prefer most remedies compared to men, except for payment, which they are less favorable towards for all four harassment types. 

\textit{Children:} Having more children is associated with higher preferences for most remedies. 

\textit{Secure:} Security in neighborhood is negatively associated with higher preference for remedies for 8 of the 28 pairings, primarily for removing content and labeling content. 

\textit{Trust:} Trust in other people is negatively associated with preferences for remedies for 19 of the 28 pairings. 

\textit{Courts:} Confidence in courts is associated with preference for the payment remedy for all harassment types but few other remedies.  

\textit{Police:} Confidence in police is not correlated with remedy preferences. 


\subsection{Qualitative responses}
We asked participants one free response question about how harassment on social media should be addressed. The most prevalent code related to the site being responsible for addressing the problem; nearly 50\% of responses referred to site responsibility, ranging from about 20\% to 60\% across countries. This was most prevalent in Malaysia, Cameroon, and the United States and least prevalent in Mongolia. Responses included content about setting policies, enforcing policy, supporting users, and protecting users. For example, a participant in Korea said: ''The social media company is primarily responsible for it, though the person who harassed others also has responsibility quite a lot.'' As part of this site responsibility, many participants described what they thought the site should do, such as: ``The social media site should give the offender a negative rating and ban them for a specific time period. This time period  being months or years or indefinitely; as well as disallowing them from creating further accounts'' (Caribbean). Some participants described why they thought social media sites were responsible, such as this one from Pakistan who said: ``This problem should be solved only by the social media website as they have all the data of the user through which they can take action against it.''

The second most prevalent code, found in nearly 25\% of responses, referred to government involvement, including regulation, police, courts, arrest, prison, criminal behavior, and juries. References to government involvement were highest in China and Pakistan and lowest in Russia, Mexico, and Cameroon. Many participants who mentioned government responsibility for online harassment indicated it should be in collaboration with law enforcement. For example, one participant from Malaysia said: ''The responsible party (social media workers) need to take this issue seriously and take swift action to `ban' the perpetrator and bring this matter to court and the police so that the perpetrator gets a fair return. This is to avoid the trauma suffered by the victim and also reduce mental illness.'' Participants varied in their indication of whether the user or the platform should report the behavior to the government. One in India said ``First of all the person who has been harassed he should simply go to police station to report this incident then he should report the account and telling his friends to report it then he should mail to instagram some screenshots of that person's chat.'' Some participants focused only on government responsibility, such as one in Colombia saying: ``Having permission for the police to see all your material on the network and electronic devices.'' 


References to managing content (e.g. removing or filtering content) and account sanctions (e.g. warnings, suspensions, bans) showed up in about 20\% of responses. These were highest in the Caribbean and the United States and lowest in China and Mongolia. Sometimes posts only recommended one step, like content removal, but more often they mentioned a multi-step process for accountability. A participant in China recommended a user rating approach: ``in serious cases, he should be banned and blacklisted, and the stains of his behavior should be recorded on his personal file.'' Many participants proposed policies that required real identities linked to the account to deter harassment and allow punishment. One person in Austria said: ``Login to the networks only with the name known to the operator. Strict consequences, first mark the content, then also delete the profile and transfer the data to the public prosecutor's office.''

About 13\% of responses referred to user responsibility, in which the person who may be experiencing or targeted by harassment should handle it themselves. These responses suggested that people should ignore harassment, stop complaining or whining about it, deal with it, and understand people will disagree. These responses were highest in Colombia and Malaysia and lowest in Austria, Cameroon, Caribbean. For example a participant in Colombia indicated that users should take steps to protect themselves: ``1. Being on social media is everyone's responsibility. 2. In social networks, you should limit yourself in admitting requests from strangers. 3. Remove and block malicious invitations.'' One in Malaysia indicated that people should work out the problems themelves: ``All parties must cast off feelings of hatred and envy towards others. To deal with this problem all parties need to be kind to each other and help each other.'' Another Malaysia participant was more explicit, saying: ``The responsible party is myself, if it happens to me, I will definitely block the profile of the person who is harassing me. The self is also responsible for not harassing others despite personal problems. It is best to complete it face to face.''

Some responses, about 7\%, referred to public awareness or public shaming as a response, which could be through  formal media coverage or offender lists or through informal user behaviors. These were highest in the Caribbean and China. One participant in Mongolia said: ``This role belongs to the private organization that runs the first social platform and to the police of the country. Disclosure to the public of the crimes committed by the perpetrators and the number of convictions related to this issue, and the damage caused by the crime.''


About 6\% of responses directly addressed the offender's role in the harassment, indicating that they are responsible and should address the problem and change their behavior. This was most prevalent in Korea and Malaysia. About 6\% referred to restitution in some way, which could involve the offender paying a fine to the site or the victim or the victim receiving compensation from the site or offender. About 6\% referred to blocking other users as a remedy.

Other codes showed up in around 5\% of responses, including verifying accounts (checking for bots, fake accounts), educating users about appropriate behaviors, and offender apologies for behavior. Apologies, when specified, were often supposed to be public rather than private, such as a participant from Cameroon's response: ``They should demand a public apology from the person, and if the person does not give the apology, the account should be banned.'' A person in China associated the public apology with reputational damage: ``Infringement of my right of reputation requires a public apology to compensate for the loss.'' In terms of account verification, some participants talked about real names or use of IPs. One from Colombia said: ``Social networks should block people's IPs and allow them to have a maximum of 2 accounts per IP, since most of the people who harass do not do it from their main accounts but rather hide inside other personalities that are not them, by Doing this would greatly reduce this type of bullying.''


\section{Discussion}
Our findings coalesce into three broad themes about global perceptions of social media harassment harms and remedies: (1) Location has a large influence on perceptions. (2) The causes are complex -- no single factor, nor even a straightforward subset of factors, emerges as a dominant predictor of perceptions of harm. (3) One-size-fits-all approaches to governance will overlook substantial differences in experiences across countries. 


\subsection{Key Role of Local Cultural Context}

Our results suggest that local cultural context plays the greatest role in determining people's perceptions of online harassment among the factors we measured. In our analysis, country emerged as the most predictive of perceptions of harm across harm types and also with respect to remedies. This is striking, especially when considering that country explained more of the variation in perceptions than gender. As is widely understood, women and girls bear a greatly disproportional brunt of harassment in general~\cite{burkett2015sex, dobson2016sext, bates2017revenge,eckert2020doxxing, barrense2020non}, and though women in each country consistently perceived greater harm than the men in the same country, women's perception of harm globally depended even more on their country. Thus for example, our data indicates that women in the United States perceive less harm from social media harassment (M=2.99, SD=1.44) than men in China (M=3.47, SD=1.2) or India (M=3.18, SD=1.33). Those are just three data points, and we do not claim that this particular set of comparisons is necessarily reliable, but it illustrates a broader point that we believe is robust from our data: \emph{some} countries' women, on average, perceive less harm under some social media harassment cases than \emph{other} countries' men, on average. 

It is unclear what it is about local cultures that has this impact (our findings suggest that there is unlikely to be a simple set of causes), and we also wish to avoid an unresolvable discussion about what exactly constitutes ``culture.'' Yet, it seems safe to conclude that a range of complex social factors that have some coherence at the national/regional level has a profound effect on how citizens of those countries and countries perceive social media harms and remedies. These are also inevitably shaped by policies and regulations in those countries. For example, some of our Malaysia participants said that online harassment should be the responsibility of the ``MCMC.'' The Malaysian Communications and Multimedia Commission is responsible for monitoring online speech, including social media, though it has little power to remove content on platforms hosted outside of Malaysia.
Though all countries we studied have some laws governing the extent of critique users can express towards their own governments, these laws vary in severity. For example, in 2015, the Malaysian government asked Facebook and YouTube to take down posts by blogger Alvin Tan which insulted Muslims \cite{bbc2015muslims}. More recently, the Indian government not only sanctioned individual users who critiqued Modi, but it sought to sanction Twitter for not taking down those posts -- Twitter has recently launched a lawsuit against the Indian government in response \cite{cnn2022twitter}. Though insults is lower in harm than other types of harassment, it is higher in some countries in our study, and it is the most prevalent type of harassment among participants in Google's 22-country survey, suggesting that it may have cumulative harmful effects for users \cite{thomas2021sok}. 

At the same time, it is important to remember that experiences and concerns within countries inevitably vary and span across boundaries. Our data indicates that reputational harm is lower in the U.S. and Austria and this may be true for the majority in our sample from those countries, but reputational harm can persist within and across boundaries. For instance, Arab women living in the U.S. may deal both with Arab and Western patriarchal structures and orientalism, thereby experiencing a form of intersectional discrimination that requires specific support measures and remedies  \cite{al2016influence}. Similarly, refugees and undocumented migrants may be less likely to report online harassment for fear of repercussion to their status in the country \cite{guberek2018keeping}. Though a focus on country-level governance is important, additional work is required to protect and support people within countries who may experience marginalization, despite or because of, local governance. 


\subsection{No Simple Causal Factors for Harm Perception}
The second broad conclusion of our study is that perceptions of harm about online harassment are complex; no simple mechanism, nor any small set of variables, easily explains relative perceptions among countries. Harm perceptions might, for example, reasonably be expected to correlate with how much people trust others, how safe they feel in their own neighborhoods, or how much they trust institutions like the police and the courts. Yet, our results find no such easy explanations: sense of neighborhood security correlated positively with greater perceptions of harm for some forms of harassment, but negatively for the nonconsensual sharing of sexual photos and sexual harassment pairing; number of children predicted greater harm for half of the harassment - harm pairings, but not the other half. 

Some correlations did emerge in our data, but it is not straightforward to interpret them. For example, trust in courts was associated with perceptions of harm in a majority of our countries. This pattern is surprising, and could indicate a desire to normalize online harassment as harmful to enable greater judicial oversight over those harms. Interestingly, trust in courts is mostly not correlated with the remedies we measured, \textit{except} for payment which is negatively correlated. It may be that lower trust in courts to procure compensation may be correlated with a higher reliance on platforms, but we would need additional data to confirm this interpretation. 


Somewhat easier to explain is that trust in other people was correlated with lower perception of harm in most cases. It may be that people who are low in trust in others assume online harassment will be severe and persistent. There was substantial variance in trust levels between countries, with Caribbean being lowest and China being highest. This suggests that harms associated with online harassment may reflect offline relationships and communities. 

Our results show that there is little or no relationship between confidence in police and harm or remedies, which may indicate that people do not see online harassment as a problem that police can or should address. This interpretation aligns also with previous research which has highlighted how police are often an inadequate organization to deal with concerns around harassment and online safety, and can sometimes cause more harm \cite{sambasivan2019they}. Instead, experts have called for investments in human rights and civil society groups who are specifically trained to support people in the communities who experience harassment \cite{york2021silicon}. Such experts could also mediate between affected people and other institutions such as the police and legal institutions. 

An exploration of factors we did not consider may find simpler or more coherent causal explanations for perceptions of harm and remedies, but we conjecture that the complexity is systemic. Online harassment, though relatively easy to discuss as a single type of phenomenon, touches on many social, cultural, political, and institutional factors, and the interplay among them is correspondingly complex. A highly patriarchal honor culture that leads women to fear the least sensitive of public exposures might be partially countered by effective law enforcement that prioritizes those women's rights; deep concerns about one's children might be offset by a high level of societal trust; close-knit communities might on the one hand provide victims with healthy support, but they might also judge and impose harsh social sanctions. 

\subsection{One Size Does Not Fit All in Online Governance}

The four types of harassment we studied all differed from each other in perceived harm, both in type of harm and severity of that harm. Non-consensual sharing of sexual photos was highest in harm, consistent with work on sexual harms that has focused on non-consensual sharing of sexual images \cite{citron2021privacy, goldberg2019nobody,dad2020most}. This work has advocated for legal protection and recourse for people who are victims of non-consensual image sharing and has brought attention to the devastating consequences it can have on victims' lives. Much of this transformative work in U.S. contexts focuses on sexual content like nude photos, which are now prohibited in some states in the U.S. (though there is no federal law) \cite{citron2019evaluating}. However, in many parts of the world there are consequences for sharing photos of women even if they do not contain nude content. 

Our findings show substantial variance in perceptions of reputational harm as well as physical harm between countries. India (medians of 4.09 and 4.01, respectively) and Colombia (4.02, 4.24) are highest in both of those categories  whereas the U.S. is lowest (2.73, 2.69). Our results corroborate Microsoft's Digital Civility Index, which found high rates of incivility in Colombia, India, and Mexico (and the U.S. being relatively low), though Russia was also high which deviates from our results. Google's survey similarly shows Colombia, India, and also Mexico as highest in prevalence of hate, abuse, and harassment \cite{thomas2021sok}. While shame associated with reputation persists globally, it may be a particularly salient factor where cultures of honor are high \cite{rodriguez2008attack}. In qualitative studies conducted in the South Asian country, including India, Pakistan, and Bangladesh, participants linked reputational harm with personal content leakage and impersonation, including non-consensual creation and sharing of sexually explicit photos \cite{sambasivan2019they}. Because women in conservative countries like India are expected to represent part of what the family considers its “honor,” reputational harm impacts not only just the individual’s personal reputation but also their family and community’s reputation.

As one South Asian activist described technology-facilitated sexual violence (quoted from \cite{maddocks2018non}):
\textit{"A lot of times, there’s an over-emphasis on sexually explicit photos. But in [this country], just the fact that somebody is photographed with another boy can lead to many problems, and we’ve seen honor killings emerging from that."} In these cases, women are expected to represent part of what the family considers its ``honor'' \cite{sambasivan2019they} and protecting this honor becomes the role of the family, and especially men in the family, who seek to regulate behavior to preserve that honor. Unfortunately, when a person becomes a victim of online abuse, it becomes irrelevant whether she is guilty or not, what matters is other people's perception of her guilt. At an extreme, families will engage in honor killings of women to preserve the honor of the family \cite{QandeelBaloch, jinsook2021resurgence, goo2019}. 

When women experience any kind of abuse, they may need to bring men with them to file reports, and then they may be mocked by officials who further shame and punish them for the abuse they experienced \cite{sambasivan2019they}. In Malaysia, legal scholars raise concern about the inadequacy of law in addressing cyberstalking in both the National Cyber Security Policy and the Sustainable Development Goals \cite{rosli2021non}. Sexual harassment, sexual harm, and reputation are strongly linked, and the threat of reputational damage empowers abusers. Many European countries have taken proactive stances against online harassment but the efficacy of their policies are not known yet. Unfortunately, any efforts to regulate content also risk threats to free-expression, such as TikTok and WeChat's suppression of LGBTQ+ topics \cite{walker2020more}. Concerns about human rights and civil rights may be especially pronounced in countries where there is not sufficient mass media interest to protest them, such as the rape of a girl in India by a high-profile politician that did not gather attention because it was outside of major cities \cite{guha2021hear}.     


In Latin American contexts, there is similar evidence that societies that place a premium on family reputation are likely to be afflicted by higher rates of intrapersonal harm \cite{osterman2011culture, dietrich2013culture}. For example, constitutional laws against domestic violence in Colombia decree that family relations are based on the equality of rights and duties for all members, and that violations are subject to imprisonment. Yet recent amendments have called for retribution against domestic violence to be levied \textit{only} when charges with more severe punishment do not apply. Human rights activists from the World Organisation Against Torture have claimed that such negligent regulations send the message that domestic violence, including harassment, is not as serious as other types \cite{omct2004violencecolombia, randall2015criminalizing}. Even with the existence of laws on domestic violence in countries like Colombia and Mexico, prevailing attitudes view harassment as a "private" matter, perhaps because of traditional norms that value family cohesion over personal autonomy. One speculation is that fears of reporting harassment because of familial backlash corroborate why survey respondents from this country may not find exposing their abusers online satisfying.

\subsection{Recommendations for Global Platform Design and Regulation}


Our recommendations for global platform design and regulation \textcolor{black}{build on work done by myriad civil society groups and follow from our own findings. In short, harms associated with online harassment is greater in non-U.S. countries and platform governance should be more actively coshaped by community leaders in those countries. }
Above all, we discourage any idea that a \emph{single} set of platform standards, features, and regulations can apply across the entire world. While a default set of standards might be necessary, the ideal would be for platforms and regulations to be further customized to local context. \textcolor{black}{A reasonable start is for platforms to regulate at the country level, though governance should be sensitive to the blurriness of geopolitical and cultures boundaries.} Digital technology is highly customizable, and it would be possible to have platform settings differ by country. Similarly, regulation of social media, as well as policy for harm caused through online interaction, should also be set locally. To a great extent the latter already happens, as applicable policy tends to be set at a national level. It should also be the case that technology companies engage with local policymakers, without assuming that one-size-fits-all approaches are sufficient. 

According to the findings discussed above, local cultural context can play an important role in helping platforms define harassment and prioritize online speech and behavior that will likely have the most impact in a given local context. For example, posting non-consensual images, whether sexual or not, can have a more severe impact in countries where women's visibility and autonomy are contentious issues. Customizing definitions of harms would also align with the task of determining the effectiveness of a remedy. If certain behaviors are criminalized offline, that would likely have an impact on how seriously platforms should take online manifestations of such harassment, and how easy it would be for users in that locality to seek help from police or courts. Lastly, due to the great variance on how local laws are shaped and implemented, platforms can play a key role in determining the effectiveness of rules as applied to them and their users. The resulting observations about what laws are effective on the ground can help platforms both customize their own policies, and engage with stakeholders more productively.

Platform features, settings, and regulation ought to be determined by multistakeholder discussions with representation from local government, local civil society, researchers, and platform creators. Input from entities familiar with the local laws, customs, and values is essential, \textcolor{black}{as others have recommended (e.g.~\cite{cammaerts2020digital,york2021silicon})}. As our study also finds, the specifics of how users respond to online harassment are localized and not given to easily generalized explanation. Of course, such discussions must be designed well. For example, we recommend that platform creators -- who have international scope yet often tend toward Western, educated, industrial, rich, and democratic (WEIRD) sensibilities~\cite{henrich2010weirdest,linxen2021weird} -- take a back seat and turn to local community leaders to lead these discussions. Platform creators have the power to determine final features anyway; additional exertion of power in such discussions will suppress local voices. Tech companies must also be willing to adopt the resulting recommendations~\cite{powell2013argument}.

Beyond platform and regulatory customization within countries, there should be transnational bodies that consider things at a global level, and which might also serve to mediate between issues that bring geographic countries into contention. Technology companies already sponsor such bodies -- for instance, Meta has a Stakeholder Engagement Team that includes policymakers, NGOs, academics, and outside experts that support the company in developing Facebook community standards and Instagram community guidelines \cite{meta2022stakeholder}. Even better would be for such bodies to have more independence, set up for autonomous governance via external organizations.

We recognize that customization by country raises new challenges, such as the question of whose policy should take precedence when cross-country interaction occurs on a platform. Or, how platforms should handle users who travel across countries (or claim to do so). Or the substantial problem, though not the focus of this paper, of how to address authoritarian regimes that are not aligned with human rights \cite{york2021silicon}. It will take work, and diplomacy, to resolve these issues, but if the aim is to prevent or mitigate harassment's harms in a locally appropriate way, the effort cannot be avoided. As to what kinds of customization such bodies might suggest, our study gestures toward features and regulations that might differ from place to place. For example, there appears to be wide variation across countries in terms of what is considered invasive disclosure. Russians generally care much less than Pakistanis whether photographs of an unmarried/unrelated man and woman are posted publicly. Thus, in some contexts, the default setting might require the explicit consent of all tagged, commented, or (automatically) recognized parties for a photo or comment to be posted. Another possibility is to adjust the ease with which a request to take down content is granted. The possibilities span a range from (A) automatically taking down any content as requested by \emph{anyone} to (Z) refusing to take down any content regardless of the volume or validity of requests. In between, there is a rich range of possibilities that could vary based on type of content and on country. With respect to how platforms manage content-removal requests, they might establish teams drawn from each geographic context, so that decision-makers address requests from cultures they are most familiar with (and based on standards recommended by the aforementioned local bodies).

\section{Conclusion}
We studied perceptions of harm and preferences for remedies associated with online harassment in 14 countries around the world. Results show that all countries perceive greater harm with online harassment compared to the U.S. and that non-consensual sharing of sexual photos is highest in harm, while insults and disrespect is lowest. In terms of remedies, participants prefer removing content and banning users compared to revealing identities and payment, though they are more positive than not about all remedies we studied. Country is the biggest predictor of ratings, with people in non-U.S. and lower income countries perceiving higher harm associated with online harassment in most cases. Most countries prefer payment, apologies, revealing identities, and rating users compared to the U.S., but are less favorable towards removing content, banning users, and labeling content. One exception to these trends is non-consensual sharing of sexual photos, which the U.S. rates more highly as sexual harassment than other countries. We discuss the importance of local contexts in governing online harassment, and emphasize that experiences cannot be easily disentangled or explained by a single factor.

\begin{acks}
This material is based upon work supported by the National Science Foundation under Grants \#1763297 and \#1552503 and by a gift from Instagram. We thank members of the Social Media Research Lab for their feedback at various stages of the project. We thank Anandita Aggarwal, Ting-Wei Chang, Chao-Yuan Cheng, Yoojin Choi, Banesa Hernandez, Kseniya Husak, Jessica Jamaica, Wafa Khan, and Nurfarihah Mirza Mustaheren for their contributions to this project. We thank Michaelanne Thomas, David Nemer, and Katy Pearce for early conversations about these ideas. 

\end{acks}

\bibliographystyle{ACM-Reference-Format}
\bibliography{main}

\appendix

\end{document}